\documentclass[aps,prl,reprint,superscriptaddress,nobibnotes,floatfix]{revtex4-2}
\usepackage{dcolumn}
\usepackage{blindtext}
\usepackage{lineno}
\usepackage{amsmath}
\usepackage{amssymb}
\usepackage{float}
\usepackage{txfonts}
\usepackage[utf8]{inputenc}
\usepackage[T1]{fontenc}
\usepackage{layouts}
\usepackage{physics}
\usepackage{graphicx}

\usepackage{hyperref}
\usepackage{braket}
\usepackage{layouts}
\usepackage{multirow}
\usepackage{subfigure}[width=0.5\linewidth]{}
\usepackage{siunitx}
\usepackage[dvipsnames]{xcolor}

\newcommand{\frep}{f_{\text{rep}}}
\newcommand{\cgain}{\mathcal{G}}

\colorlet{lcu_color}{NavyBlue}
\hypersetup{
	linkcolor  = lcu_color,
	citecolor  = lcu_color,
	urlcolor   = lcu_color,
	colorlinks = true,
	breaklinks = true
}

\begin{document}

\title{Electric-field metrology of a terahertz frequency comb using Rydberg atoms}

\author{Wiktor Krokosz}
\thanks{These authors contributed equally}
\email{w.krokosz@cent.uw.edu.pl}
\affiliation{Centre for Quantum Optical Technologies, Centre of New Technologies, University of Warsaw, Banacha 2c, 02-097 Warsaw, Poland.}
\affiliation{Faculty of Physics, University of Warsaw, Pasteura 5, 02-093 Warsaw, Poland.}

\author{Jan Nowosielski}
\thanks{These authors contributed equally}
\affiliation{Centre for Quantum Optical Technologies, Centre of New Technologies, University of Warsaw, Banacha 2c, 02-097 Warsaw, Poland.}
\affiliation{Faculty of Physics, University of Warsaw, Pasteura 5, 02-093 Warsaw, Poland.}

\author{Bartosz Kasza}
\affiliation{Centre for Quantum Optical Technologies, Centre of New Technologies, University of Warsaw, Banacha 2c, 02-097 Warsaw, Poland.}
\affiliation{Faculty of Physics, University of Warsaw, Pasteura 5, 02-093 Warsaw, Poland.}

\author{Sebastian Borówka}
\affiliation{Centre for Quantum Optical Technologies, Centre of New Technologies, University of Warsaw, Banacha 2c, 02-097 Warsaw, Poland.}
\affiliation{Faculty of Physics, University of Warsaw, Pasteura 5, 02-093 Warsaw, Poland.}

\author{Mateusz Mazelanik}
\affiliation{Centre for Quantum Optical Technologies, Centre of New Technologies, University of Warsaw, Banacha 2c, 02-097 Warsaw, Poland.}

\author{Wojciech Wasilewski}
\affiliation{Centre for Quantum Optical Technologies, Centre of New Technologies, University of Warsaw, Banacha 2c, 02-097 Warsaw, Poland.}
\affiliation{Faculty of Physics, University of Warsaw, Pasteura 5, 02-093 Warsaw, Poland.}

\author{Michał Parniak}
\email{michal.parniak@uw.edu.pl}
\affiliation{Centre for Quantum Optical Technologies, Centre of New Technologies, University of Warsaw, Banacha 2c, 02-097 Warsaw, Poland.}
\affiliation{Faculty of Physics, University of Warsaw, Pasteura 5, 02-093 Warsaw, Poland.}





\begin{abstract} 
Terahertz radiation finds an increasing number of applications, yet efficient generation and detection remain a challenge and an active area of research.  In particular, the precise detection of weak and narrowband terahertz signals is notoriously difficult. Here, we employ a novel type of single-photon detector based on Rydberg atoms to both detect and calibrate a terahertz frequency comb over an octave-spanning range, yet with a MHz-level selectivity. We calibrate the intensity of the electric field of the comb against the fundamental atomic properties, while achieving the intensity (power) sensitivity down to \SI{45.2}{\femto\watt\per\cm\squared\per\Hz\tothe{1/2}} (\SI{1.84}{\femto\watt\per\Hz\tothe{1/2}}) within a single mode of the frequency comb, all in a room-temperature operated setup. Our results elucidate the transition of terahertz frequency combs into the quantum regime, enabling high-precision and high-sensitivity spectroscopy. This breakthrough allows terahertz science to better leverage revolutionary techniques developed for optical frequency combs.
\end{abstract}
\maketitle 
\section{Introduction}
THz (Terahertz) and millimeter-wave radiation occupy the spectral region between the microwave and infrared domains. Historically, this range was one of the least explored parts of the electromagnetic spectrum due to the lack of efficient sources and detectors. However, rapid progress in photonics, semiconductor technologies, and materials science has brought THz technologies to the forefront of modern research. Today, THz radiation is finding growing applications in diverse areas, such as high-speed wireless communications \cite{Liu_2024_2}, security imaging, non-destructive testing, spectroscopy \cite{Liebermeister_2021,Couture_2023}, and advanced material characterization, e.g., study of exciton dynamics and phase transitions in semiconductors \cite{Chen_2007,Venanzi_2024}.

Despite this progress, key challenges remain, particularly in generating and detecting coherent THz signals with high sensitivity and resolution. Pulsed THz sources that generate electromagnetic frequency combs have emerged as powerful tools for addressing these challenges. Frequency combs—well known in the optical domain—enable high-precision spectroscopy and frequency metrology and are now being extended into the THz range \cite{Yasui_2006, Mouret_2009, Yasui_2010, Kocybik_2024, Preu_2011, Finneran_2015, Shin_2023}. Developments in THz comb generation include techniques based on photomixers driven by continuous-wave lasers, passive mode-locking \cite{Hiraoka_2022}, and referencing to optical frequency combs for improved accuracy. These combs have enabled hyperspectral imaging \cite{Sterczewski_2019}, phase-resolved measurements \cite{Arikawa_2024}, and studies of phase noise in THz systems \cite{Krause_2024}.

On the detection side, significant progress has been made in fabricating fast, multi-pixel THz detectors, yet fundamental limits in sensitivity—especially at low photon flux—continue to hinder applications in quantum sensing and metrology. This has motivated the exploration of unconventional detection platforms, such as atomic vapor-based sensors.

In particular, Rydberg-atom ensembles have emerged as promising candidates for THz detection due to their extreme sensitivity to electric fields, stemming from their large polarizabilities and strong dipole transitions. These sensors rely on well-understood principles of atomic physics and offer a range of functionalities: fluorescence-based imaging of THz and mmWave fields \cite{Holloway_2014,Wade_2016,Wade_2018,Downes_2020,Chen_2022,Downes_2022,Downes_2023}, direct field reception \cite{Gordon_2014,Legaie_2024}, and even THz generation schemes via nonlinear optical processes in Rydberg media \cite{Lam_2021}. The versatility of these systems has also been demonstrated in demodulation of broadband signals \cite{Meyer_2023}, Doppler-free spectroscopy using multi-photon transitions \cite{Bohaichuk_2023}, and long-distance sensing in open-air environments \cite{Otto_2023}. Polarization-insensitive detection schemes \cite{Cloutman_2024}, access to high-angular-momentum states \cite{Allinson_2024}, and applications such as mmWave thermometry in cold-atom and ambient settings \cite{Schlossberger_2025} have further expanded the scope of Rydberg sensors. Moreover, recent efforts have explored their utility in the calibration of commercial radar systems \cite{Borowka_2024_2}. A comprehensive review of these developments can be found in \cite{Schlossberger_2024_2}.

In this paper, we report the observation and analysis of a THz frequency comb using a Rydberg-atom-based detector. Our experimental approach combines two key methods: Autler-Townes splitting measurements, which allow for precise determination of the THz electric field strength in absolute units \cite{Sedlacek_2012}, and a microwave-to-optical conversion technique \cite{Borowka_2023}, which extends the sensitivity down to the level of thermal radiation. Using these tools, we characterize the frequency comb by identifying the amplitudes and absolute mode numbers of its individual spectral components.

Furthermore, we explore the interaction of the frequency comb with different Rydberg transitions, providing both experimental measurements and comparisons with theoretical models. The conversion scheme enables us to probe a broad spectral range and observe the structure of the comb around various atomic resonances. We also present direct measurements of beat-note signals between adjacent comb lines, revealing visible high-frequency components and enabling insight into the underlying coherence and stability of the comb structure.

Our results demonstrate the potential of Rydberg-atom detectors in THz frequency metrology and highlight their applicability in emerging quantum-enabled sensing technologies.

\section{Experimental setup}

\subsection{Detection}

In this work, a detection setup in two configurations is employed: the first one is the standard A-T (Autler-Townes) splitting measurement used for calibration to SI units, while the second is an improvement over microwave-to-optical conversion, enabling ultrasensitive detection of higher RF (radio frequencies).

The first auxiliary method relies upon a $^{87}$Rb vapor cell and two optical fields. The laser beams are introduced into the cell from opposing directions
and overlaid colinearly. This setup is the
typical two-photon excitation scheme utilized in Rydberg atom electrometry,
where usually two lasers are used to excite the atoms. The \SI{780}{\nm} probe beam
tuned to the D2 line is scanned in frequency and, after interaction with atomic vapors is shone upon an avalanche
photodiode (Thorlabs APD430A) to obtain the probe transmission spectrum. By the observation of the EIT (electromagnetically induced transparency) effect, introduced by
the counter-propagating \SI{483}{\nm} coupling laser tuned to the $5^{2}\text{P}_{3/2}(\text{F}=3)\rightarrow 27^{2}\text{D}_{5 / 2}$ transition, the setup is capable of performing RF electric field measurements by analyzing
the Autler-Townes splitting in the domain of the probe laser detuning \cite{Sedlacek_2012}.

\begin{figure*}[htbp]
    \centering
    \includegraphics[width=\linewidth]{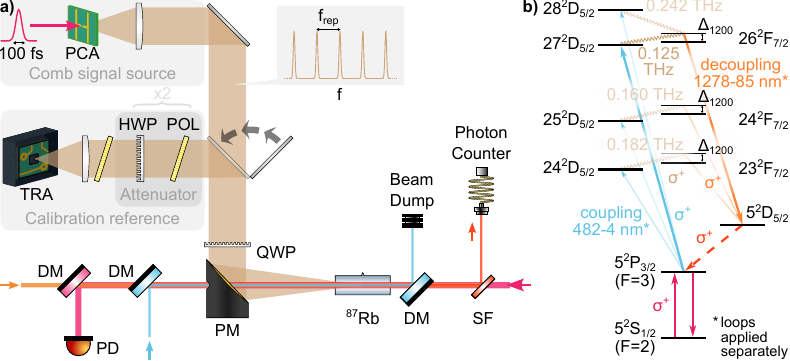}
    \caption{\textbf{a)} Simplified scheme of the setup used for the calibration and detection of THz frequency comb signal. $\mathsf{TRA}$ -- automotive radar chip in an RF absorbing box, $\mathsf{HWP}$
    -- half-waveplate, $\mathsf{QWP}$ -- quarter-waveplate, $\mathsf{PD}$ -- photodiode,
    $\mathsf{PM}$ -- parabolic mirror, $\mathsf{PCA}$ -- photoconductive antenna,
    $\mathsf{POL}$ -- polarizer, $\mathsf{DM}$ -- dichroic mirror, $\mathsf{SF}$ --
    spectral filter. The removable metallic mirror allows switching between signal sources. The TRA beam has slanted polarizers to discard reflections, while the HWP + POL setup has two copies to increase the achievable attenuation.
    \textbf{b)} State-level diagram of fields and state transitions facilitating
    the conversion of frequencies centered around various detected frequencies to the 776 nm output
    optical signal. The targeted THz signal frequencies are \SI{0.125}{\THz}, \SI{0.160}{\THz}, \SI{0.182}{\THz} and \SI{0.242}{\THz}. Note that only one THz frequency is targeted at a time -- the coupling and decoupling lasers are coarsely detuned to address the relevant transition. Furthermore, by fine-detuning the decoupling laser by $\Delta_{1200}$ it is possible to tune to the frequencies near the chosen transition.}
    \label{fig:setup}
\end{figure*}

This measurement scheme has a fundamental limitation on the minimal resolvable
field, resulting from the finite EIT linewidth, caused by the movement of
atoms at room temperature and the arising Doppler broadening. With the objective of measuring much weaker fields, the
A-T setup is then extended by introducing a third laser beam, namely
the 1278 nm decoupling field tuned to the $26^{2}\text{F}_{7/2}\rightarrow 5^{2}\text{D}_{5/2}(\text{F}=4)$ transition, realizing the full setup, visible in Fig.~\ref{fig:setup}~\textbf{a)}.

This constitutes the second and main measurement method, utilizing parametric RF-to-optical upconversion via six-wave mixing \cite{Borowka_2023}.
By transferring the atomic population down from the Rydberg state, it allows for
a radiative, 776 nm transition between $5^{2}\text{D}_{5/2}(\text{F}=4)$ and $5^2\text{P}_{3/2}(\text{F}=3)$ states
to take place, resulting in the emission of optical photons, driven by the incident
THz fields.
The conversion output beam is then separated using a dichroic mirror and a spectral filter,
before being sent to a single photon counter in the form of SPAPD (single photon
avalanche photodiode), Excelitas SPCM-AQRH. Finally, the SPAPD photon counts are registered with an externally triggerable IDQ 900 time controller.

Such a detection scheme achieves superior detection sensitivity to the intensity of THz fields,
down to the thermal noise limit, with the potential for time-resolved
measurements and synchronization with a pulsed source. Thanks to having a strong reference field source, it is possible to
calibrate the photon count readout to the electric field amplitude measured
from A-T splitting. What is more, as is the case with A-T
splitting, we are not limited to resonant RF detection and addressing different
frequencies is possible by detuning the coupling or decoupling fields. For
simplicity, this work explores tuning just the decoupling laser for mapping the
bandwidth of a particular RF transition.

Both detection methods can be carried out in a consecutive manner, with minimal effort required for switching between them. The coupling and decoupling laser frequencies are stabilized to a narrowband frequency-doubled fiber laser (NKT Photonics, \SI{1560}{\nm}), which is itself stabilized to a Rb cell via a modulation-transfer lock. The conversion scheme requires that the probe laser is also stabilized to the frequency-doubled laser via an optical phase-loop, which also drives the frequency scans in A-T measurements. The beams within the cell are collimated to have the matching Gaussian waists of \SI{250}{\um} with probe, coupling and decoupling beam powers of \SI{140}{\uW}, \SI{600}{\mW}, and \SI{126}{\mW} respectively. To optimize the conversion process, the vapor cell is heated to \SI{50}{\degreeCelsius} using a custom made heater utilizing hot air to avoid interference with metal elements.

\subsection{Signal generation}

The THz field is introduced from a high-power source, namely an automotive radar chip TRA\_120\_045 from Indie Semiconductor. While A-T measurements may be performed with the signal entering the vapor cell from the side, the parametric conversion necessitates phase-matching, so the THz field is combined colinearly with optical fields and focused using a parabolic mirror with a through hole for optical beams. This enables colinear propagation of all fields in the vapor cell, and the setup is further improved using custom-made metamaterials, which allow for beam collimation, attenuation, and polarization control \cite{Borowka_2024_2}. Both lenses and waveplates are made using an FDM (fused deposition modeling) 3D printer and HIPS (high impact polystyrene) filament, while the polarizers are custom-made flex PCBs (printed circuit boards). All of the designs are fine-tuned for the operating transition frequency using FDTD (finite difference time domain) simulations. As indicated in the Fig.~\ref{fig:setup}, the TRA source is enclosed in an RF absorbing box with a small aperture facing the first lens. The enclosure pre-attenuates the source, minimizes reflections from it and ensures that only the radiation passing through the aperture couples to the atoms.

As such, a sufficiently powerful source, like the aforementioned radar chip, is
employed for calibration and connecting the results obtained in both detection methods. Subsequently, we can switch to conversion detection
and study lower intensity field sources, in particular, a THz frequency comb.
We therefore introduce a PCA (photoconductive antenna -- BATOP iPCA-21-05-300-800-h), which, given a bias voltage, emits THz pulses following the intensity characteristic
of a femtosecond Spectra-Physics MaiTai laser, which has a repetition rate
$\frep$ of $\SI{79.49}{\MHz}$. The PCA has a hyperhemispherical silicon lens
mounted on top, which partially focuses the output. It is then collimated and
sent into the setup with the same path, and a metal sheet is used as a mirror
to toggle between the high-power chip and the locally weaker PCA.

All of the fields, including the THz beams, reach the vapor cell with circular polarizations, which drive transitions between the atomic levels of the highest possible total angular
momenta, exhibiting larger dipole moments.

\section{Results}

\subsection{Calibration}

\subsubsection{A-T splitting calibration}

Due to its low intensity, the THz frequency comb can be observed only via the conversion detection, and as such, its strength can be expressed only in relative units of photon counts per second. However, by performing the absolute calibration of the receiver and tying together A-T splitting and photon counts regimes, the strength of the low intensity field, down to thermal level, can be depicted in terms of electric field amplitude.

We begin the calibration by measuring the A-T splitting in the probe transmission
spectrum. For this measurement, the decoupling laser is covered, and only the effects
from the coupling laser and THz field can be seen. The probe transmission
spectrum is then measured for attenuations of the THz field between 0 and \SI{1.5}{\dB}
introduced by two attenuators, each consisting of a half-waveplate and a polarizer. The A-T splitting,
created by the THz field, can then be directly observed in the probe transmission spectra as the splitting of the EIT caused by the \SI{483}{\nm} laser. Next, we determine the Rabi frequencies $\Omega$
for the corresponding splitting by fitting the numerical solutions of the steady state to the measured probe transmission spectra \cite{Kasza2024}.
Subsequently, knowing the dipole moment $d$ of the
considered THz transition, the strength of the incoming field can be expressed
in the amplitude of the electric field $E$ by the relation
\begin{equation}
    E = \frac{h}{|d|}\Omega,
\end{equation}
where $h$ is Planck's constant.
In the experiment, the dipole moment of the main considered atomic transition, $27^2\text{D}_{5/2}\rightarrow26^2\text{F}_{7/2}$, is equal to
$d = 594a_0 e$ \cite{Sibalic_2017}, where $a_0$ is the Bohr radius and $e$ is
the value of the elemental charge. The calibration of the A-T splitting regime is then determined by fitting a linear function with a fixed slope, equal to $-1$, to the calculated electric field amplitudes as a function of THz field attenuation.

\begin{figure*}[htbp]
    \centering
    \includegraphics[width=\linewidth]{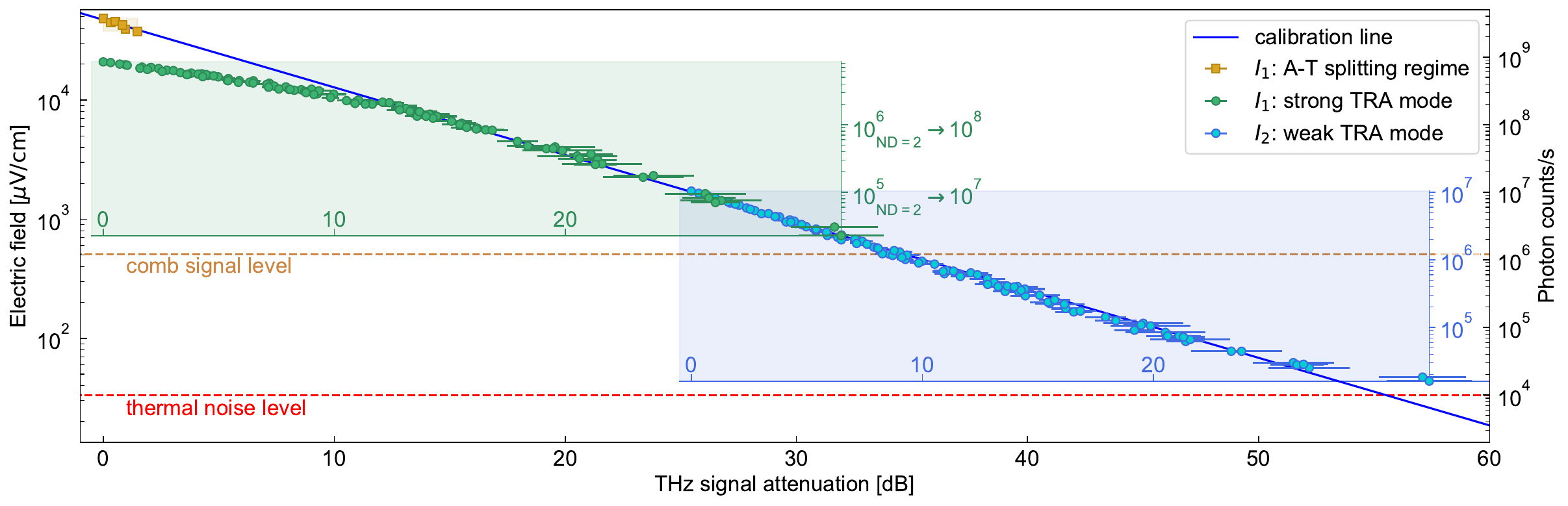}
    \caption{Calibration curve derived for the $27^{2}\text{D}_{5/2}\rightarrow 26^{2}\text{F}_{7/2}$ THz transition from the single photons and A-T splitting
    measurements, together with the fitted calibration line. and both the comb tooth signal
    and the thermal noise levels. Colored areas correspond to the measurement series for the TRA source powers $I_1$ (green, yellow) and $I_2$ (blue). Different plot markers represent A-T splitting (square) and photon counts measurement (circle). Each measurement series involved continuously attenuating the same signal using the same set of HWPs and polarizers. For the attenuation values represented on the horizontal axis, the photon counts from the SPAPD were registered and are displayed on the right vertical axis, while the A-T splitting measurements yielded the electric field values visible on the left one. The strong TRA mode measurements had to be performed with a neutral density (ND) filter put before the photon counter, to avoid the regime where detector dead times influence the results. The higher THz powers deviate from the calibration line due to the atomic saturation of the conversion process, while the much slower deviations in the high attenuation regime are caused by the imperfections of the used HWPs and polarizers. For reference, the dashed lines show the measured photon counts of the converted frequency comb signal and thermal radiation.}
    \label{fig:calibration}
\end{figure*}

\subsubsection{Photon counts calibration}

To perform the measurements of converted photon counts, the probe laser is locked at the transition resonant frequency, the decoupling laser is turned on, and both the coupling and decoupling lasers are tuned to maximize the conversion signal coming from the comb tooth closest to the resonance. Then, the photon count measurements of the converted signal are performed for the attenuation range spanning from 0 to 32 dB via the setup of HWPs and polarizers.
To further extend the calibration range, the measurements are done for two TRA source power levels, which we will label as: $I_1$ used for the A-T splitting measurements, and $I_2$, for which the A-T splitting cannot be measured in the probe transmission spectrum. The $I_1$ power according to the manufacturer's documentation is equal to $I_1 = \SI{-7}{\text{dBm}}$. However, the real power entering the setup is significantly lower due to the pre-attenuation caused by the small aperture in the enclosure. As the conversion signal due to the $I_1$ power saturates the SPAPD, for those measurement series, an additional neutral density (ND) filter is added before the photon counter.  Finally, we turn off the TRA chip and measure the photon counts corresponding to the frequency comb tooth closest to the atomic resonance. Additionally, we cover both THz sources and measure the conversion signal corresponding to the thermal radiation, and then cover the decoupling laser, which allows for the measurement of the dark counts.

Similarly to the calibration of the A-T splitting regime, the line function with a fixed slope is fitted to the data for both source power levels. Points lying near the thermal noise and saturation level, defined as a point where the response of the receiver becomes nonlinear, are omitted to increase the fit accuracy. Hence, the linear function is fitted only to points lying between \SI{11,5}{\dB} (saturation level) and \SI{45}{\dB} (near the thermal level) THz signal attenuation. Both of those threshold points have been determined by finding the point for which additional omission of data points does not impact the fit accuracy. As both measurements are performed for the same attenuation range, 
the ratio $I_2/I_1$ can be 
determined precisely using fitted intercept line parameters, with their values being
$I_2/I_1 = \SI{25,48 \pm 0,07}{\dB}$ which gives $I_2 = \SI{-32,48 \pm 0,07}{\text{dBm}}$.
This way, a full response curve spanning from atomic saturation to thermal radiation level is created. 

Finally, the scaling factor between the electric field amplitude and measured photon counts is calculated by using the fitted intercept line parameters
yielding the absolute calibration of the receiver.
The measured points, scaled using the aforementioned steps 
together with the absolute calibration line, the thermal noise, and the comb tooth level, are shown
in Fig.~\ref{fig:calibration}, with the square and circle plot markers representing the A-T splitting and photon counts measurement accordingly. The colored regions shown in the figure correspond to the measurement regions for both TRA source powers, with vertical axes representing the number of measured photon counts per second in both cases, and horizontal axes representing the THz attenuation values in both measurements. Additionally, the noticeably small amount of measured data points for the A-T splitting measurement can be attributed to the comparably low sensitivity of this measurement technique.
Based on the absolute calibration, we then find that the electric field amplitude of the measured THz frequency comb tooth is equal
to $E_\text{c}= \SI{405\pm8}{\micro \V \per \cm}$ and the amplitude of thermal noise integrated in the whole conversion bandwidth is equal
to $E_\text{noise}=\SI{36,7\pm0,7}{\micro \V \per \cm}$. The calculated values, as well as the entire calibration, are determined without subtracting the dark counts of the detector. It can also be noted based on Fig.~\ref{fig:calibration} that the response to the frequency comb tooth lies in the detector linear response regime, which we define as a linear dependence of the conversion signal on the power of the oncoming THz field. Finally, to extend the context of the measured values, we may derive rough estimates of the total power of the measured fields. Assuming that the THz field ($\lambda = \SI{2.38}{\milli\meter}$) is focused down to a diffraction-limited area $A=\pi(1.22\lambda N)^2\approx 27\ \mathrm{mm}^2$ by the parabolic mirror with f-number $N=1$ we get a lower-bounded power for the pre-attenuated TRA (corresponding to $E_{\mathrm{TRA}}=\SI{5e4}{\micro \V \per \cm}$) $P_{\mathrm{TRA}}\approx-30\ \mathrm{dBm}$ and for a single tooth of the comb $P_{c}\approx-72.3\ \mathrm{dBm}$.

\subsection{Tunable broadband readout of frequency comb signal}

\begin{figure*}[htbp]
    \centering
    \includegraphics[width=\linewidth]{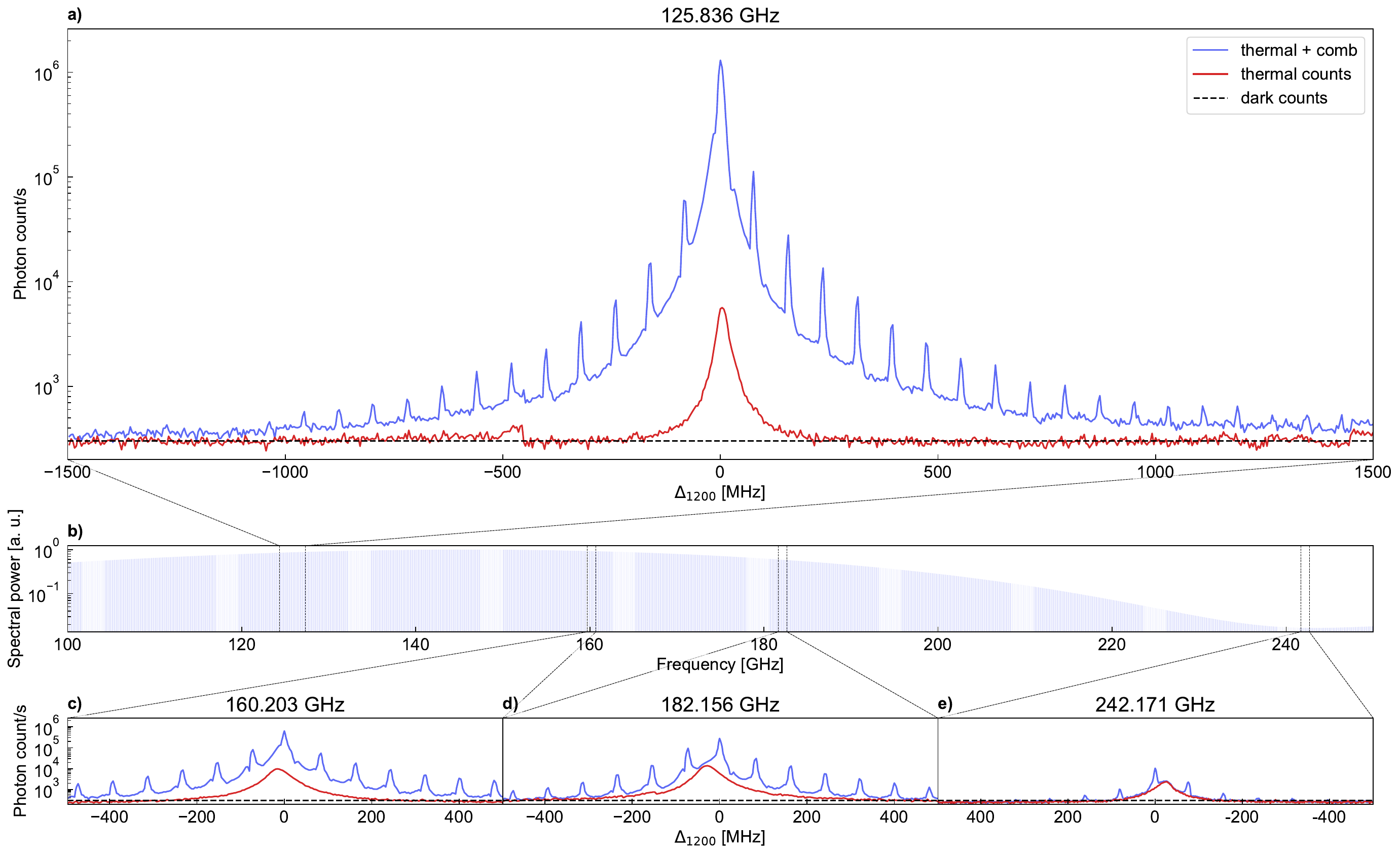}
    \caption{\textbf{a)} Measured number of converted photons per second as a function
    of decoupling laser detuning from the $27^{2}\text{D}_{5/2}\rightarrow 26^{2}\text{F}_{7/2}$ transition.
    \textbf{b)} Spectral  power distribution of the frequency comb, obtained using time-domain
    spectroscopy.
    \textbf{c)}-\textbf{e)} Measured number of converted photons per second as
    a function of the decoupling laser detuning in range of \SI{1}{\GHz} around the
    $25^{2}\text{\text{D}}_{5/2}\rightarrow 24^{2}\text{F}_{7/2}$,
    $24^{2}\text{D}_{5/2}\rightarrow 23^{2}\text{F}_{7/2}$, and $26^{2}\text{D}_{5/2}\rightarrow 28^{2}
    \text{F}_{7/2}$ atomic transitions respectively. Note that Subfigure \textbf{e)} corresponds to a downward transition, as can be seen in Fig.~\ref{fig:setup}. As such, the decoupling laser needs to be detuned in the opposing direction to properly offset the THz detuning.} 
    \label{fig:scan}
\end{figure*}

Next, we visualize the THz frequency comb by scanning the decoupling
laser, thus tuning the detector to different teeth (frequency modes) of the comb. During the
scanning process, both probe and coupling lasers are locked at the optimal
frequencies. For the considered transition, we perform a scan over the range of \SI{3}{\GHz}
with a step of \SI{4}{\MHz} around the atomic resonance. The measurement is
performed by detuning the decoupling laser over the frequency range
mentioned above and measuring the number of converted photons per second for
each of the detunings. As can be seen in Fig.~\ref{fig:scan}~\textbf{a)}, the distance
between two subsequent peaks is the same as the MaiTai repetition rate $\frep$
and remains constant across the scan range, with a regime of higher than unity SNR
(signal-to-noise ratio) spanning over more than 2 GHz. Additionally, scans
without the comb are performed to observe the intensity of converted thermal radiation.

This bandwidth corresponds to a single energy
transition. We are, however, able to map different regions of the frequency comb
spectrum. We tune both the coupling and the decoupling lasers to different
atomic transitions, allowing us to detect the THz comb around different frequencies.
Using the ARC library, several candidates have been selected and can be seen in
Fig.~\ref{fig:scan}~\textbf{c)}-\textbf{e)}. Namely, we detect the THz field at the frequencies of
$\SI{160}{\GHz}$, $\SI{182}{\GHz}$, and $\SI{242}{\GHz}$, corresponding to the
$25^{2}\text{D}_{5/2}\rightarrow 24^{2}\text{F}_{7/2}$,
$24^{2}\text{D}_{5/2}\rightarrow 23^{2}\text{F}_{7/2}$, and $26^{2}\text{D}_{5/2}\rightarrow 28^{2}\text{F}
_{7/2}$ atomic transitions respectively. All of them are depicted in Fig.~\ref{fig:setup}~\textbf{b)}. Similarly to the previous scan, we perform
the measurements by detuning the decoupling laser every \SI{4}{\MHz} step over the
frequency range of \SI{1}{\GHz}, and measure the photon counts per second for each
step. Noticeably, the maximum converted signal drops for higher THz frequencies.
We attribute this behavior to the strength of the PCA antenna being dependent on the emitted frequency. Using a standard
THz detection scheme of time-domain spectroscopy, the spectral shape of an
individual pulse is measured to verify the spectral power distribution,
visible in Fig.~\ref{fig:scan}~\textbf{b)}. While the resolution is limited to $\sim\SI{10}
{GHz}$, it confirms an order of magnitude difference in the signal
strength. The lower dipole moments also contribute to lower measured photon counts, however the variations are not significant, ranging from $\SI{594}{a_0 e}$ for \SI{125}{\GHz} to $\SI{353}{a_0 e}$ for \SI{242}{\GHz} as calculated by ARC \cite{Sibalic_2017}. 
 
\begin{figure}[htbp]
    \centering
    \includegraphics[width=\linewidth]{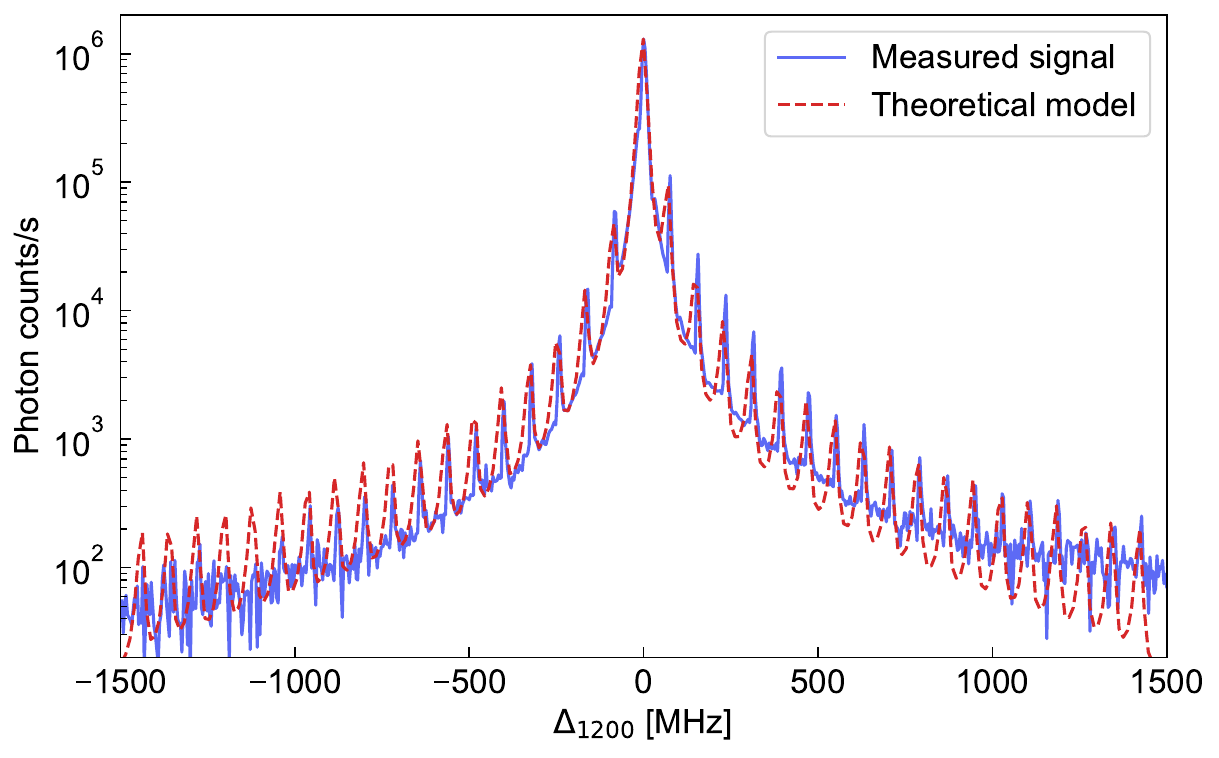}
    \caption{The tunable broadband readout of frequency comb around \SI{125}{\GHz} performed by the scan of decoupling laser. The thermal photon signal is subtracted and we fit the theoretical prediction of the comb under the assumption of the detector linear response regime. }
    \label{fig:skladak_fit}
\end{figure}

The behavior of the comb frequency scans shown in Fig.~\ref{fig:scan} can be further understood by comparing them with the numerical predictions. Under the assumption of the linear response, justified by the absolute calibration of the receiver, the expected photon counts as a function of the decoupling laser detuning $\Delta_{1200}$, for the measurement time much longer than $\frep$ is given as:
\begin{equation}
    \mathcal{N}_\text{comb}(\Delta_{1200}) \propto \sum_n I_n|\cgain(\Delta_\text{THz}+nf_\text{rep}, \Delta_{1200})|^2, \label{eq:comb = sum THz detunings}
\end{equation}
where $I_n$ is the intensity of each tooth and $\Delta_{\text{THz}}$ corresponds to the frequency distance of the closest tooth from the transition resonance. The conversion gain $\cgain$ is the ratio between the amplitude of the THz and converted fields \cite{Andrews_2014, Rueda_2016}, in this case proportional to the atomic coherence between $5^2 \text{D}_{5/2}$ and $5^2 \text{P}_{3/2}(\text{F}=3)$ states. As the rest of the lasers remain at the constant frequency, we only consider the dependence of $\cgain$ on the detuning of the THz field and the decoupling laser. Based on the linear assumption, we then fit the theoretical model to the measured comb frequency scan shown in Fig.~\ref{fig:scan}~\textbf{a)} with the thermal photons subtracted. The theoretical model consists of the steady-state solutions for each of the teeth considered, which are then combined according to Eq.~\eqref{eq:comb = sum THz detunings} to recover the full scan. The measured comb scan with the fitted theoretical predictions can be seen in Fig.~\ref{fig:skladak_fit}. It can be noted that the theoretical model and its envelope match the measurements well, proving the applicability of the detector's linear response regime. The divergence of the model, visible for the higher decoupling laser detunings, is a result of the influence of other excited states, namely the hyperfine structure of the $5^2\text{D}_{5/2}$ state, which has not been taken into account in the simulations.

\subsection{Deriving the sensitivity and thermal power}

In order to assess the robustness of the THz sensor and perform a quantitative comparison with other platforms, we evaluate its sensitivity and noise equivalent power (NEP). The measured thermal radiation includes other noise sources, such as photon counter dark counts, and as such is considered as the noise floor.

The derivations are performed for the main transition under consideration, i.e., the $27^{2}\text{D}_{5/2}\rightarrow 26^{2}\text{F}_{7/2}$ transition at \SI{125.836}{\giga\hertz}. We measure the system noise with photon counts, using acquisition times of $t = \SI{1}{\second}$, and measuring for \SI{30}{\s} in total. We get the mean counts of $\mathcal{N}_\text{noise} = \SI{9.87(25)d3}{}$ photons per second. We denote the standard deviation of this statistic as $\sigma_\text{noise} = \SI{250}{}$ photons per second. Noticeably, the statistics are super-Poissonian (for Poissonian $\sigma_\text{noise} = \sqrt{\mathcal{N}_\text{noise}} = \SI{99}{}$ photons per second should follow), which is expected for thermal light, but for single shot measurements of multimode light at a measurement time far exceeding the bandwidth inverse, the quantum phenomena of thermal bunching should cancel out. What is instead observed are the macroscopic and classical gain fluctuations of the setup, which can be calculated to constitute a relative error of about \SI{1.5}{\percent}. Shorter measurements are more likely to approach the shot noise (Poissonian) limit, as well as mitigate these macro effects; however, this is not a priority of this work.

Following the calibration stage, presented in Fig.~\ref{fig:calibration}, the converted photon counts can be translated into the equivalent incident electric field of $E_\text{noise} = \SI[]{36.7(7)}{\micro\volt\per\centi\meter}$ and further into intensity through $I_\text{noise} = E_\text{noise}^2 / 2 Z_0 = \SI[]{1.78(7)}{\pico\watt\per\centi\meter\squared}$, where $Z_0 \approx \SI{377}{\text{Ohm}}$ is the impedance of free space. Next, to obtain NEP, the receiver's effective aperture needs to be considered. For this, we can consider the gain-defined small antenna aperture $A$ as:
\begin{equation}
    A = g \frac{\lambda^2}{4\pi},
\end{equation}
where $g$ is the antenna gain (directivity) and $\lambda = \SI{2.38}{\milli\meter}$ is the wavelength of received field. We obtain the antenna gain $g=9$ from field phase-matching considerations (i.e., the receiver reception pattern) \cite{Borowka_2023}, which leads to the result of $A = \SI{4.07}{\milli\meter^2}$. Consequently, the effective registered power corresponding to the thermal radiation equal to $P_\text{noise} = I_\text{noise} A = \SI{72.5(27)}{\femto\watt}$ can be retrieved.

Finally, the parameters of interest can be calculated. Power sensitivity $\xi$ is derived as the thermal intensity $I_\text{noise}$ scaled by the ratio of thermal photon count fluctuations $\sigma_\text{noise}$ to the mean counts $n_\text{noise}$, and acquisition time $t$, and calculated as $\xi = I_\text{noise} \frac{\sigma_\text{noise}}{\mathcal{N}_\text{noise} \sqrt{t}} = \SI[]{45.2(17)}{\femto\watt\per\centi\meter\squared\per\hertz\tothe{0.5}}$. Consequently, NEP equals to this sensitivity multiplied by $A$, yielding a value of $\text{NEP} = \xi A = \SI[]{1.84(7)}{\femto\watt\per\hertz\tothe{0.5}}$. Note that the aperture used in the above calculation is significantly smaller than the aperture used for estimating the power of the TRA and comb sources, and thus the derived NEP value corresponds to the best case scenario - i.e. perfect matching of the source's radiation mode to the sensor. The full values and uncertainties considered in this derivation are presented in Tab.~\ref{tab:thermal} in Appendix E.

\subsection{Deriving the frequency comb mode number}

Energy state levels are usually calculated using the ARC library, which, in the case of the evaluated RF transitions in rubidium-87, relies on experimental data of finite precision. On top of that, there are numerous additional effects, such as Stark or Zeeman shifts, which become even more cumbersome with temperature-dependent optical density variations. Therefore we propose an alternative method to estimate the detected frequency by determining the mode (tooth) number of frequency comb.

Observing particular teeth and performing frequency scans happens through
detuning from the resonant case, which happens in relative units. By
estimating the mode number, we can move to absolute frequencies,
due to the information embedded in the comb peak separation. Knowing that the mode
number is an integer, it is possible to infer the transition frequency, which, for some transitions, can be further verified using the radar chip. As it is
stabilized using a phase-locked loop, it has a known frequency and can confirm
the validity of the derivations.

We have determined the MaiTai repetition rate $\frep$ to be $\SI{79,49020(1)}{\MHz}$ using Agilent N9010A spectrum analyzer. In
the case of the main transition under consideration, an ARC value of \SI{125,836(4)}{\GHz} is calculated, with the uncertainty obtained by redoing the calculations and propagating the uncertainty of quantum defects \cite{Han_2006, Wenhui_2003}. The next step involves locating the working point, at which the scan visible in Fig.~\ref{fig:scan}~\textbf{a)} is performed. The thermal profile is equivalent to an integral over the THz frequencies, so by fitting a Lorentzian profile, the resonant $\Delta_{1200}$ value can be found. This particular model shows excellent agreement with the measured thermal photon data, indicating that the resonance is at $\Delta_{1200} = \SI{4,92(5)}{\MHz}$.

The theoretical considerations, presented with the broadband scan in Fig.~\ref{fig:skladak_fit} are then used to yield the position of the tooth closest to the atomic resonance. It is located at the detuning of \SI{-3,213(39)}{\MHz}, which in absolute units gives \SI{125,828(4)}{\GHz}. Dividing by $\frep$ results in \SI{1582,934(64)}{} passing the $3\sigma$ test with respect to the nearest integer. Given the aforementioned lack of consideration for the shifts in energy levels, this result gives a strong indication that the mode number of this particular tooth is equal to 1583. Reverting the above derivations, taking the optical ruler as the ground truth, we arrive at a transition frequency of \SI{125,8411(32)}{\GHz}. 

Given the small but still relevant deviation from the value provided by ARC, an additional verification is performed using the radar chip source. As it is controlled by a phase-locked loop with a known reference frequency, we are able to assess the frequency of the comb tooth to be located at a target frequency of \SI{125,829(1)}{\GHz} corresponding to a mode number estimate of \SI{1582,95(4)}{}, confirming the previous result.
The parameters used in the derivation of tooth mode number are presented in Tab.~\ref{tab:ordinal}.

\begin{table}[H]
    \centering
    \caption{Numerical parameters and their uncertainties, used to obtain the frequency comb mode number, tabularized for clarity.}
    \begin{tabular}{lr}
        \hline
        Parameter name                & Value and uncertainty            \\
        \hline
        ARC transition frequency      & \SI[]{125.836(4)}{\giga\hertz}   \\
        Repetition rate               & \SI[]{79.4902(20)}{\mega\hertz}  \\
        Thermal peak center           & \SI[]{4.92(5)}{\mega\hertz}      \\
        Main comb peak center         & \SI[]{-3.21(4)}{\mega\hertz}     \\
        Reference frequency           & \SI[]{125.831(4)}{\giga\hertz}   \\
        Main comb peak frequency      & \SI[]{125.828(4)}{\giga\hertz}   \\
        Mode number                   & \SI[]{1582.93(6)}{}              \\
        Measured transition frequency & \SI[]{125.8411(32)}{\giga\hertz} \\
        \hline
    \end{tabular}
    \label{tab:ordinal}
\end{table}

\subsection{Beating of adjacent frequency comb peaks}

Further evidence of frequency comb detection can be presented by measuring
the beat notes between individual peaks. As they are separated by multiples of
$\frep$, the beat note frequencies will also assume values of $n\frep$. The beating
manifests itself in the time domain, so we observe the time-resolved
photon conversion response on the timetagger, connected to the single photon counter.
The 40 MHz MaiTai internal photodiode readout is used as the external trigger of the timetagger, allowing for measurement of all the potential beat-note frequencies.

As with the frequency scans, detuning the decoupling laser allows adjusting
the sensitivity to different frequencies and a value that sits between two
comb peaks are selected so as to make the beatings more pronounced in comparison to the resonant case.

\begin{figure}[htbp]
    \centering
    \includegraphics[width=\linewidth]{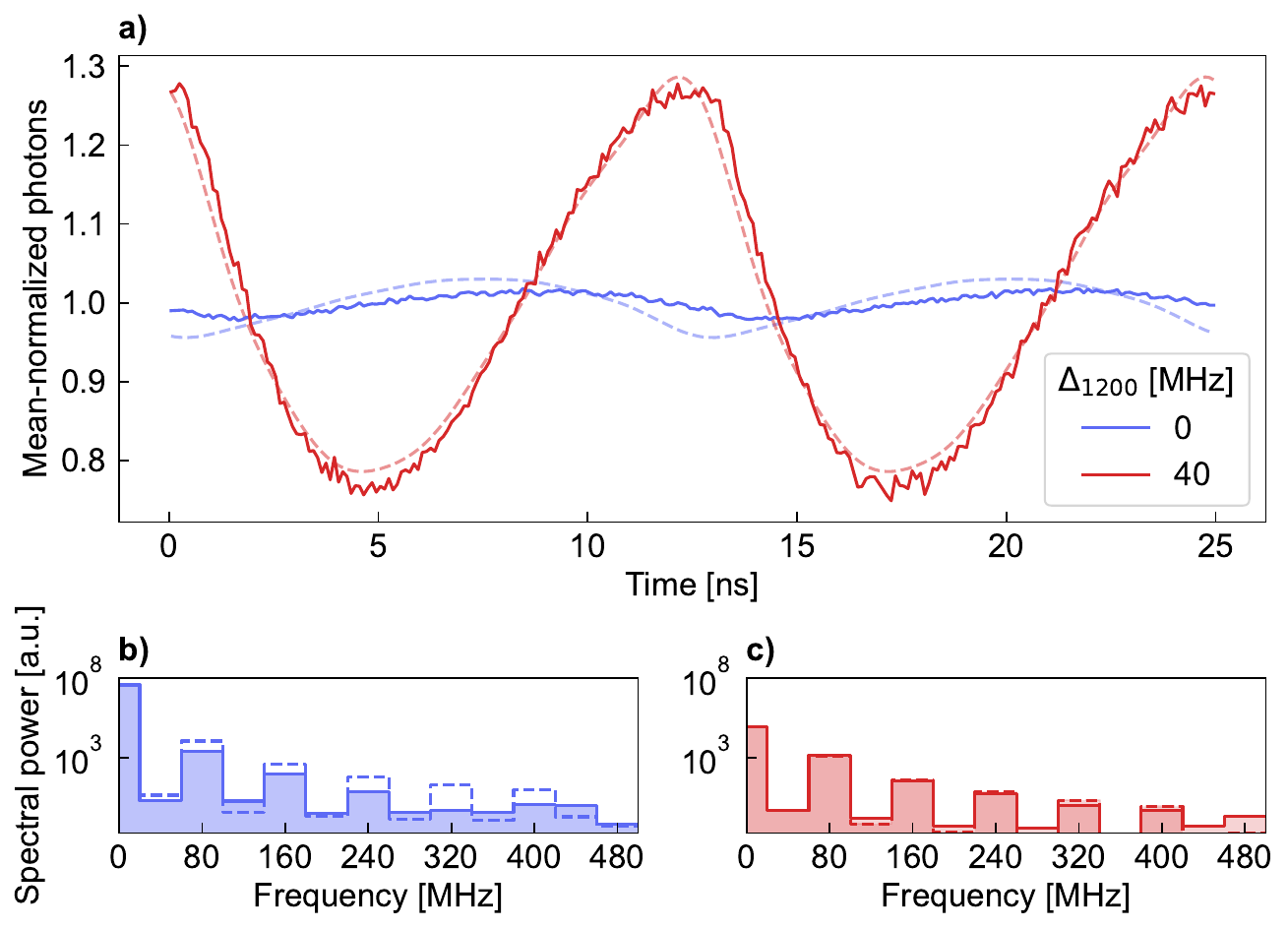}
    \caption{\textbf{a)} Temporal profile of the measured conversion
    signal, corresponding to mean-normalized photon counts. The blue lines correspond
    to the frequency comb resonant peak condition, while the case for red lines is detuned by
    $\Delta_{1200} = \SI{40}{\MHz}$. Dashed lines represent results obtained from applying Eq.~\eqref{eq:beat_final} to the fitted conversion gain. \textbf{b)}-\textbf{c)} The
    Fourier transforms of the respective temporal profiles, with the resonant case having significantly
    lower contrast between the oscillating terms and the constant contribution. Again, dashes represent theoretical predictions, with the signal amplitudes scaled to experimental data levels. Contrary to \textbf{a)}, the full signal is presented here.}
    \label{fig:dud}
\end{figure}

The photons are counted for \SI{60}{\s}. The observed signals, as seen
in Fig.~\ref{fig:dud} are modulated mainly with the first-order beating contribution, but performing a Fourier transform reveals subsequent multiples of $\frep$ as
well. As detuning the decoupling laser decreases the conversion efficiency, the collected signals are mean-normalized and presented in Fig.~\ref{fig:dud}~\textbf{a)}.
Positioning our detector frequency sensitivity at a detuning corresponding to $\frep
/ 2$ corresponds to significantly more pronounced modulation, clearly
visible in the temporal signal, while the Fourier transform in Fig.~\ref{fig:dud}~\textbf{b)}-\textbf{c)}
reveals the existence of higher-order beat frequencies. It can also be noted that the spectra differ visibly in the contribution of the constant, non-oscillating term, with the spectrum for the detuned case having significantly higher contrast of the oscillating terms. This observation is consistent with the comparison of the signals in the time domain shown in Fig.~\ref{fig:dud}~\textbf{a)}.

The dynamics of the beating of adjacent comb teeth can be understood by once again investigating the function describing the expected converted signal, which we write this time as:
\begin{equation}
\mathcal{N}_{\text{comb}}(t) \propto \left|
\sum_n A_n e^{-i\omega_n t} 
\cgain(\Delta_\text{THz} + nf_\text{rep}, \Delta_{1200})
\right|^2,
\end{equation}
where $A_n$ is the amplitude of the $n$-th frequency component, and $\omega_n = 2\pi n \frep$. For the convenience of notation, we denote the conversion gain $\cgain(\Delta_\text{THz}+nf_\text{rep}, \Delta_{1200})$ as $\cgain_n$. The equation above can be split into two parts, namely the non-oscillating part, described by Eq.~\eqref{eq:comb = sum THz detunings}, and the time-dependent part given as:
\begin{equation}
    \mathcal{N}_{\text{beat}}(t) \propto \sum_{n\neq m}A_nA^*_m e^{-i\delta_{nm}t}\cgain_n\cgain^*_m,
    \label{eq:beat_v1}
\end{equation}
with $\delta_{mn} = 2\pi(n-m)\frep$ being the beating frequency between $m$-th and $n$-th comb components. It is worth noting that for the signal averaged over a time longer than the period of the slowest oscillating term, all the oscillating terms vanish, leading to the expected converted signal given by Eq.~\eqref{eq:comb = sum THz detunings}. The terms visible can be further grouped by the value of $\delta_{nm}$:
\begin{equation}
    \mathcal{N}_{\text{beat}}(t) \propto \sum e^{-i2\pi n\frep t} \sum_{m}A_{m+n}A^*_m \cgain_{m+n}\cgain^*_m.
    \label{eq:beat_final}
\end{equation}

As can be seen in the equation above, the beating only occurs at frequencies that are multiples of $\frep$. Moreover, as all of the terms in the mentioned equation are time-dependent, they do not contribute to the constant term of the measured beat note signal, and thus the difference in the beating contrast can be attributed mostly to the part of the conversion signal described by Eq.~\eqref{eq:comb = sum THz detunings}. 

\section{Discussion}
In this paper, we have demonstrated the detection of a THz frequency comb using a microwave-to-optical converter based around a warm Rydberg atomic ensemble. We further characterized the frequency comb by deriving its electric field amplitude and performing wide frequency scans of the conversion signals. Finally, we measured the beating between adjacent frequency components. We have also verified the linear behavior of the converter in the wide frequency range by comparing the frequency scans and beat note signal to the theoretical model.

The obtained THz intensity detection NEP values show an improvement of 2 to 4 orders of magnitude over state of the art room temperature alternatives \cite{Schuetz_2005, Qi_2014, Kim_2021, Xi_2025, Wang_2025, Mateos_2020, Bandurin_2018}, showcasing great potential for Rydberg atoms as an alternative platform in this field. Similar conclusions can be drawn from other attempts at using a Rydberg atom-based converter for THz detection \cite{Li_2024}, where we also achieved an improvement of about 4 orders of magnitude in sensitivity. Direct comparison of NEP is difficult in this case, due to different assumptions in aperture derivation.

The are multiple issues to be solved that are beyond the scope of this work. Namely, due to the THz frequency comb being weak, determining its electric field amplitude requires an additional, stronger source. This issue, in theory, can be overcome by increasing the sensitivity of the A-T splitting detection, which can be achieved by considering the Doppler canceling scheme \cite{Bohaichuk_2023}, leading to a narrower EIT peak and the possibility of direct measurement of the A-T splitting caused by the frequency comb.

We believe that our results pave the road for future applications of Rydberg atom-based detectors in the rapidly growing field of THz electrometry. The unique possibility of determining the strength of weak fields in absolute units, as well as overall high sensitivity, makes Rydberg atom-based setups a great candidate for ultra-sensitive and easy-to-calibrate THz detectors. Additionally, the setup can be easily extended to perform multiple tasks, such as THz imaging or spectroscopy.

\section*{Acknowledgments}
We thank M.~Lipka for his initial contribution to the construction of the pulsed THz source. This research was funded in whole or in part by the National Science Centre, Poland, grants No.~2021/43/D/ST2/03114 and No.~2024/53/N/ST7/02730. The "Quantum Optical Technologies" (FENG.02.01-IP.05-0017/23) project is carried out within the Measure 2.1 International Research Agendas programme of the Foundation for Polish Science, co-financed by the European Union under the European Funds for Smart Economy 2021-2027 (FENG). S.B.~is a recipient of the  Foundation for Polish Science START 2025 scholarship.

\section*{Disclosures}
The authors declare no conflict of interest.

\section*{Data availability}
Data underlying the results presented in this paper are available in the Ref.~\cite{rydcomb_data}.

\section*{Code availability}
The codes used for the numerical simulation are available from W.K.~upon request.

\bibliography{references}

\section*{Appendix A: Calibration of the half-waveplates' transmission}
\begin{figure}[htbp]
    \centering
    \includegraphics[width=\columnwidth]{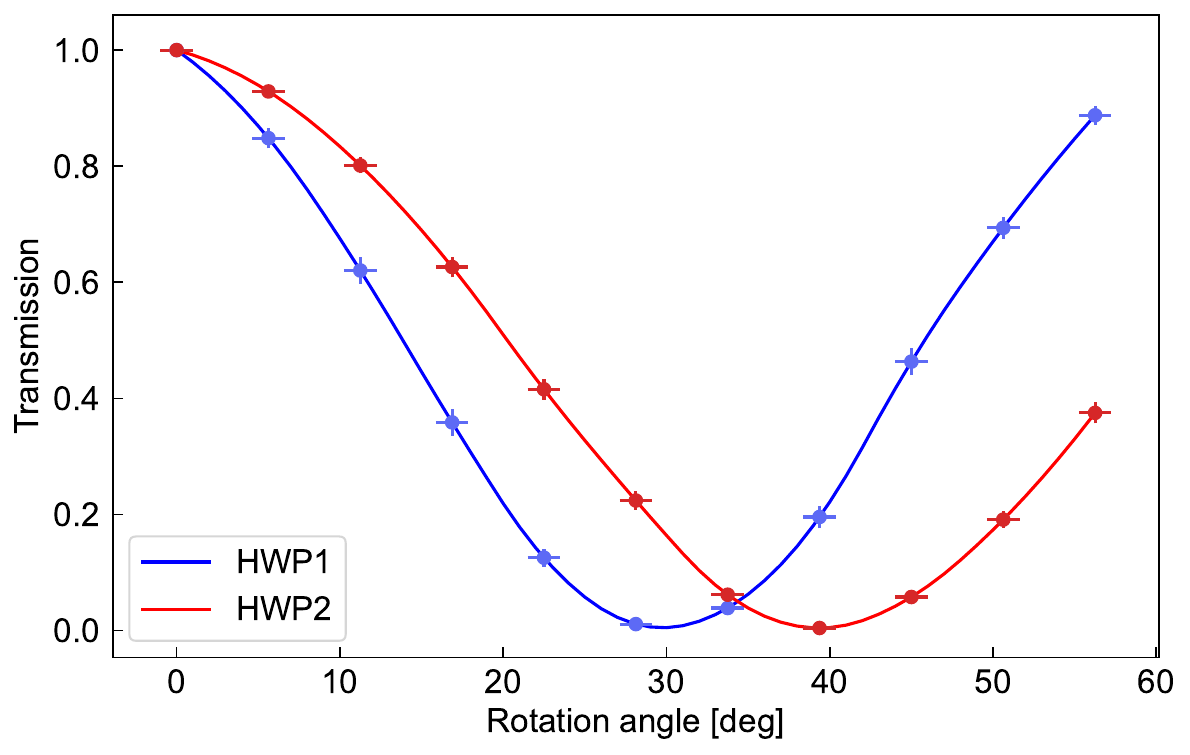}
    \caption{Measured transmission of both polarizer and half-waveplate pairs as a function of the waveplate rotation angle, together with interpolated calibration functions. HWP1 and HWP2 correspond to the first and second pair.}
    \label{fig:hwp}
\end{figure}

To determine the attenuation imposed by the HWP combined with the polarizer, the transmission of such a pair needs to be calibrated as a function of the HWP rotation angle, for which we use the photon counts measurement as a function of angles of both HWP for the $I_2$ TRA source intensity. For the sake of calibration of each of the individual waveplates, we choose the cross-sections of measured photon counts through the rotation angle of the other waveplate corresponding to its measured maximum transmission. As the HWPs are not fully rigid and could move by about $\pm 0.5^\circ$ on their own, the uncertainty of the transmission functions is taken as a difference between the interpolated function and the interpolated function shifted by $0.5^\circ$. The interpolated transmission functions together with the data points and uncertainties can be seen in Fig.~\ref{fig:hwp}. Additionally, based on the found transmission functions, maximum attenuations of both pairs can be estimated as $A_1 = \SI{19.6\pm1.7}{\dB}$ and $A_2 = \SI{23,6\pm0,4}{\dB}$ for the first and second pair, respectively.

\section*{Appendix B: Theoretical model} \label{appendix: Theoretical model}

The experimental setup consists of a five-level atomic open quantum system, which is described by the GKSL (Gorini–Kossakowski–Sudarshan–Lindblad) equation:
\begin{equation}
\begin{split}
   \dot{\rho}(t) = &-\frac{i}{\hbar}\left[H, \rho(t)\right] + \mathcal{L}_\Gamma \left[\rho(t) \right], 
\label{eq: gksl}
\end{split}
\end{equation} 
where $H$ is the Hamiltonian of the system, $\rho$ is the density matrix, and $\mathcal{L}_\Gamma$ is a Lindblad superoperator, which accounts for decoherence mechanisms. We model a five-level atomic ladder in the RWA (rotating-wave approximation), which produces the following form of the Hamiltonian in dipolar approximation:
\begin{equation}
H = -\tfrac{\hbar}{2}
\begin{pmatrix}
\scriptstyle 0 & \scriptstyle \Omega^*_{780} & 0 & 0 & 0 \\
\scriptstyle \Omega_{780} & \scriptstyle i\Gamma_1 + 2\Delta_\textbf{I} & \scriptstyle \Omega^*_{480} & 0 & 0 \\
0 & \scriptstyle \Omega_{480} & \scriptstyle i\Gamma_2 + 2\Delta_{\textbf{II}} & \scriptstyle \Omega^*_{\text{THz}} & 0 \\
0 & 0 & \scriptstyle \Omega_{\text{THz}} & \scriptstyle i\Gamma_3 + 2\Delta_{\textbf{III}} & \scriptstyle \Omega^*_{1200} \\
0 & 0 & 0 & \scriptstyle \Omega_{1200} & \scriptstyle i\Gamma_4 + 2\Delta_{\textbf{IV}}
\end{pmatrix} - \tfrac{i \hbar \Gamma_\text{tr}}{2}\mathbb{I},
\end{equation}
where $\Omega_{ij}$ are Rabi frequencies driving transitions between states $\ket{i}\rightarrow\ket{j}$, $\Gamma_i$ is the decay rate of state $\ket{i}$ and $\Gamma_\text{tr}$ is a parameter modeling the effective broadening coming from limited interaction time of atoms with thin laser beams. Finally, the detunings of the incident electric fields from the atomic resonances are grouped into common terms:
\begin{equation}
    \begin{split}
        \Delta_{\textbf{I}} &= \Delta_{780}, \\
        \Delta_{\textbf{II}}&= \Delta_{780} + \Delta_{480}, \\
        \Delta_{\textbf{III}} &= \Delta_{780} + \Delta_{480} + \Delta_{\text{THz}}, \\
        \Delta_{\textbf{IV}} &= 
        \Delta_{780} + \Delta_{480} + \Delta_{\text{THz}} - \Delta_{1200}.
    \end{split}
\end{equation}

Room temperature atomic vapors are significantly influenced by the Doppler effect, stemming from the atoms moving along the electric field propagation axis. Factoring in the broadening of the resonance can be simply implemented as additional diagonal values of the Hamiltonian. For that, we define the $K$ matrix, whose diagonal elements represent cumulative sums of the wave vectors $k$ of the fields required to excite the system to the corresponding energy levels, with each subsequent level involving an additional driving field. The signs are picked to reflect the counter propagation of the probe beam, yielding:
\begin{equation}
K = \begin{pmatrix}
0 & 0 & 0 & 0 & 0 \\
0 & \scriptstyle -k_{780} & 0 & 0 & 0 \\
0 & 0 & \scriptstyle -k_{780} + k_{480} & 0 & 0 \\
0 & 0 & 0 & \scriptstyle -k_{780} + k_{480} + k_{\text{THz}} & 0  \\
0 & 0 & 0 & 0 & \scriptstyle -k_{780} + k_{480} + k_{\text{THz}} + k_{1200}
\end{pmatrix}.
\end{equation}
Then we can write our full Hamiltonian as:
\begin{equation}
    H_v = H + K v,
\end{equation}
where $v$ denotes the velocity class of atoms,
and obtain the final solution by averaging over the Maxwell-Boltzmann velocity distribution, represented by a statistical weight of $w(v)$:
\begin{equation}
    \rho = \int w(v) \rho(v) d v.
\end{equation}
The obtained Hamiltonian is then evaluated numerically.

\subsection*{Modeling of Autler-Townes splitting}
The Autler–Townes effect is observed as a splitting and broadening of the EIT in the absorption spectrum of the probe laser. Probe laser absorption is proportional to expected values of $\bra{1}\rho(\Delta_{780})\ket{0}$. In the experiment, the measurement of absorption happens after some time from the moment the interaction started, which allows the system to relax into a steady state. In the macroscopic picture, the steady state is time independent, $\dot \rho = 0$, which simplifies the calculations significantly. 

\subsection*{Modelling of terahertz frequency comb}
On the other hand, for the analysis of the conversion signal caused by the THz frequency comb, we focused mostly on the conversion gain $\cgain$ and the conversion signal $\mathcal{N}$. In light of the model described above, the conversion signal is given as:
\begin{equation}
    \mathcal{N} \propto \left|\bra{1}\rho\ket{4}\right|^2.
\end{equation}

To better understand the conversion gain, it is worth first considering the conversion of the monochromatic THz signal. Under the assumption of the detector's linear response, the conversion gain $\cgain$ can be interpreted as the ratio between the emitted and THz fields, usually expressed in terms of their intensities as:
\begin{equation}
    |\cgain|^2 = \frac{\mathcal{N}}{I_\text{THz}}.
\end{equation}
It can be noted that due to the THz field being monochromatic, the value of $I_\text{THz}$ is just a scaling factor. Thus, the conversion gain becomes proportional to the conversion signal, and the following relation can be established:
\begin{equation}
    \mathcal{N} \propto |\cgain|^2 \propto |\bra{1}\rho\ket{4}|^2.
\end{equation}

\section*{Appendix C: Determining Rabi frequency from A-T splitting}

\begin{figure}[htb]
    \centering
    \includegraphics[width=\columnwidth]{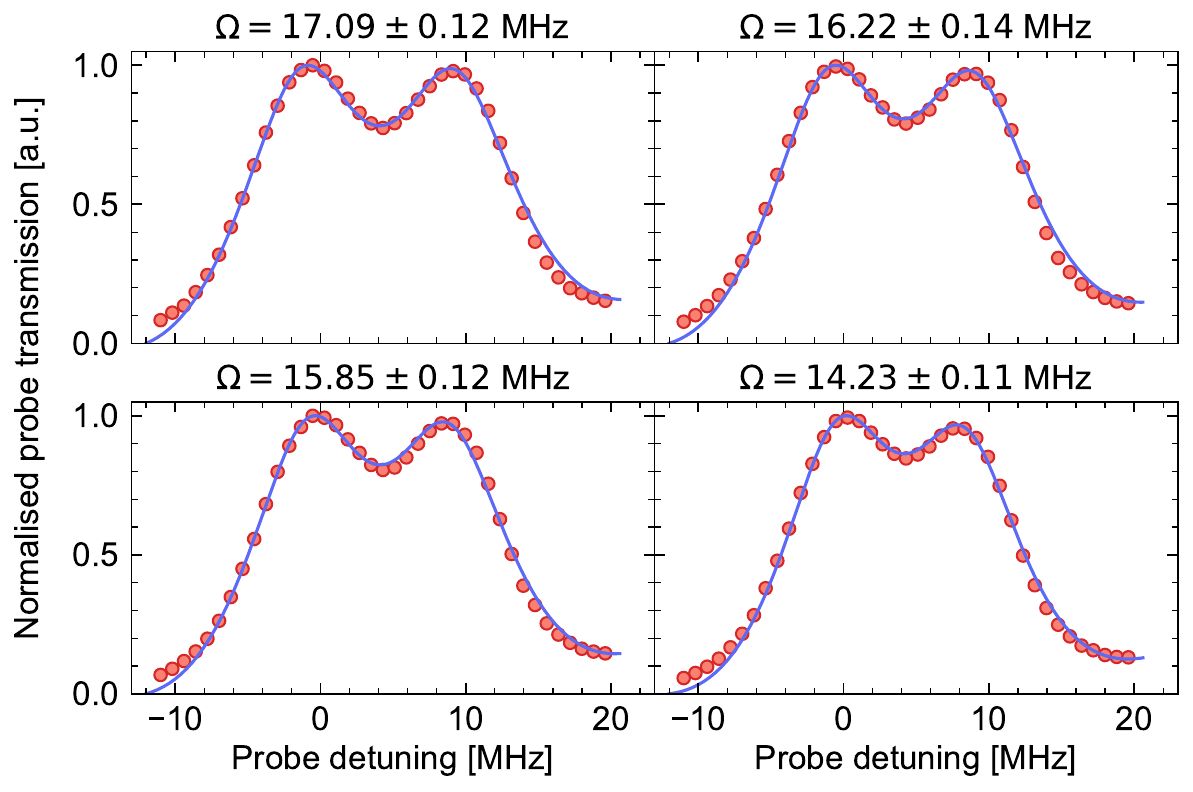}
    \caption{Probe transmission spectra of A-T splitting obtained for different intensities of THz field generated by the TRA in the calibration procedure. The results are presented together with the numerical simulations of the steady state. Both the experimental and numerical data are min-max normalized.}
    \label{fig:A-T}
\end{figure}

As mentioned in the main text, for the sake of the calibration of the receiver, the Rabi frequency of the THz field is determined based on the measured A-T splitting visible in the probe transmission spectra. The values of the Rabi frequencies are found by numerically fitting the steady-state solution for the theoretical model described in Appendix B to the probe transmission signal. In such a fit, the Rabi frequency of the THz field is treated as one of the parameters of the fitted function. Exemplary probe transmission spectra, together with the fitted theoretical model, can be seen in Fig.~\ref{fig:A-T}. In order to perform the fitting properly, both the experimental and numerical data are min-max normalized. To further increase the accuracy of the numerical fits, the common parameters, which remained constant during the measurement, such as the Rabi frequencies of laser fields, are fitted simultaneously to all the measured spectra.

\section*{Appendix D: Selective observation of comb teeth} \label{appendix: maps}

In order to get a better intuition of the non-standard measurements performed in this paper, we have used the theoretical framework and experimental data to produce a plot of the conversion gain $\cgain\left(\Delta_\text{THz}, \Delta_{1200}\right)$ values. It is important to note that the derivation of this value is phenomenological in its nature, as it relies on the min-max normalization of the experimental data and outputs of the theoretical model. The fitting procedure on normalized detuning scans is then performed to retrieve the parameters found in the system's Hamiltonian, which is subsequently used to calculate the value of $\cgain$ for arbitrary detunings.

\begin{figure}[htbp]
    \centering
    \includegraphics[width=\columnwidth]{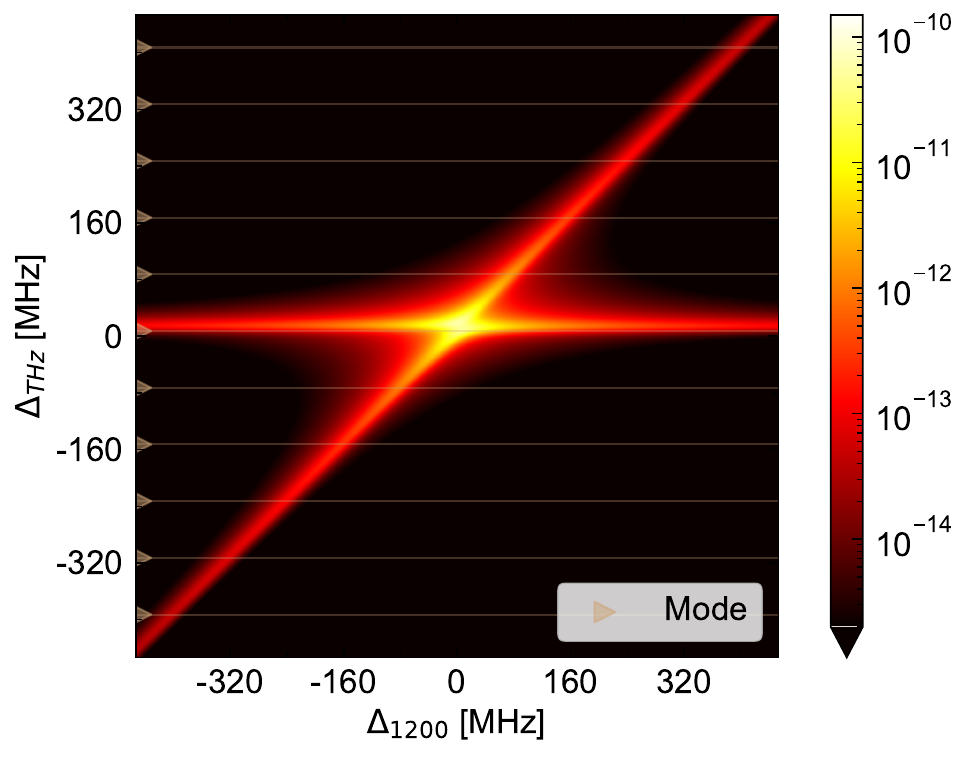}
    \caption{Map of $|\cgain|^2$ with respect to $\Delta_\text{THz}$ and $\Delta_{1200}$. It is directly proportional to the conversion signal intensity, and as such, the color map is trimmed at the order of magnitude corresponding to the noise floor. The comb teeth are marked on the vertical axis and can be thought of as sampling points for a given value of $\Delta_{1200}$.}
    \label{fig:scan_map}
\end{figure}

The calculated conversion gain can be seen in Fig.~\ref{fig:scan_map}, showing two resonances that give a good working intuition for performing comb spectroscopy with this setup. A single photon count measurement, given a frequency comb and a fixed value of $\Delta_{1200}$, can be understood as a discrete sum of conversion gain values along a vertical line of the map, as expressed in Eq.~\eqref{eq:comb = sum THz detunings}. The frequency comb teeth remain static when tuning the decoupling laser, meaning that we adjust the conversion gain to the positions of the teeth. The resonances observed in Fig.~\ref{fig:scan} are spaced at $\frep$, because we count two teeth at this interval. On the other hand, the thermal photon measurements correspond to taking an integral across the $y$-axis.

\begin{figure}[htbp]
    \centering
    \includegraphics[width=\columnwidth]{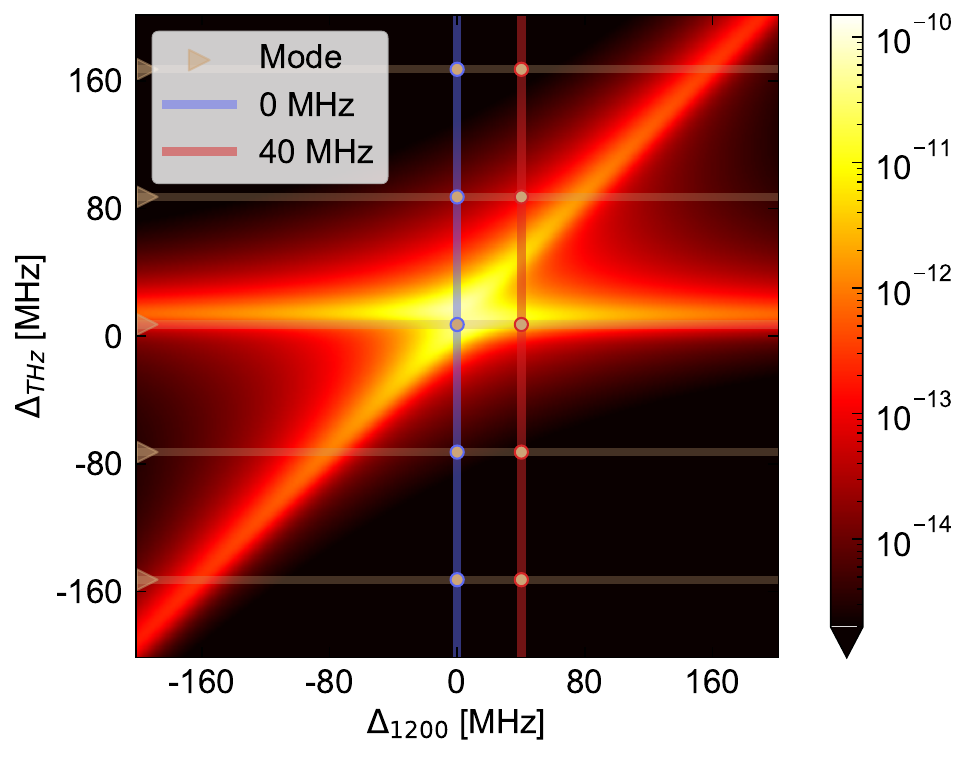}
    \caption{Zoom of the $|\cgain|^2$ map from Fig.~\ref{fig:scan_map}. Here, the particular detuning values of $\Delta_{1200}$ chosen for evaluating the beating of adjacent frequency comb teeth are displayed with their respective colors. The vertical lines correspond to the decoupling laser detunings analyzed in Fig.~\ref{fig:dud}. Contributions to the conversion output signals are then marked by the intersections with frequency comb modes.}
    \label{fig:dud_map}
\end{figure}

The zoomed-in map in Fig.~\ref{fig:dud_map} showcases the asymmetric profile of the conversion gain that could be observed in the performed scans.

\section*{Appendix E: Sensor's thermal power parameters} \label{appendix: Thermal parameters}

The sensor's parameters, used in the considerations of thermal noise level, as well as the sensor's sensitivity and NEP, are presented in Tab.~\ref{tab:thermal}.

\begin{table}[H]
    \caption{Numerical results with their uncertainties obtained when deriving the sensitivity and noise equivalent power (NEP) of the detector, tabularized for clarity.}
    \centering
    \begin{tabular}{lr}
        \hline
        Parameter name         & Value and uncertainty                                                     \\
        \hline
        Transition frequency   & \SI[]{125.836(4)}{\giga\hertz}                                            \\
        Transition wavelength  & \SI[]{2.38241(8)}{\milli\meter}                                           \\
        Small antenna aperture & \SI[]{4.06504(26)}{\milli\meter\squared}                                  \\
        Thermal photon counts  & \SI[]{9.87(25)d3}{\mathrm{phot}/\mathrm{s}}                                                       \\
        Thermal field          & \SI[]{36.7(7)}{\micro\volt\per\centi\meter}                               \\
        Thermal intensity      & \SI[]{1.78(7)}{\pico\watt\per\centi\meter\squared}                       \\
        Thermal power          & \SI[]{72.5(27)}{\femto\watt}                                                \\
        Sensitivity            & \SI[]{45.2(17)}{\femto\watt\per\centi\meter\squared\per\hertz\tothe{0.5}} \\
        NEP                    & \SI[]{1.84(7)}{\femto\watt\per\hertz\tothe{0.5}}                          \\
        \hline
    \end{tabular}
    \label{tab:thermal}
\end{table}

\end{document}